\documentclass[10pt,twocolumn,letterpaper]{article}

\usepackage[pagenumbers]{wacv} % To force page numbers, e.g. for an arXiv version

\usepackage{pifont}   % \ding{51}/\ding{55} check and cross marks in tables
\usepackage{booktabs}
\usepackage{amsmath,amssymb}
\usepackage{multirow}

\definecolor{wacvblue}{rgb}{0.21,0.49,0.74}
\usepackage[pagebackref,breaklinks,colorlinks,allcolors=wacvblue]{hyperref}

\def\wacvPaperID{2813} % *** Enter the WACV Paper ID here
\def\confName{WACV}
\def\confYear{2027}

\newcommand{\AccBbbBright}{1.000}
\newcommand{\AccBbbClean}{1.000}
\newcommand{\AccBbbCrfA}{1.000}
\newcommand{\AccBbbCrfB}{1.000}
\newcommand{\AccBbbCrfC}{0.978}
\newcommand{\AccBbbCropA}{0.523}
\newcommand{\AccBbbCropB}{0.811}
\newcommand{\AccBbbFps}{1.000}
\newcommand{\AccBbbNoise}{1.000}
\newcommand{\AccBbbRot}{0.735}
\newcommand{\AccBbbScale}{1.000}
\newcommand{\AccJellyBright}{0.998}
\newcommand{\AccJellyClean}{0.996}
\newcommand{\AccJellyCrfA}{0.996}
\newcommand{\AccJellyCrfB}{0.990}
\newcommand{\AccJellyCrfC}{0.952}
\newcommand{\AccJellyCropA}{0.499}
\newcommand{\AccJellyCropB}{0.531}
\newcommand{\AccJellyFps}{0.997}
\newcommand{\AccJellyNoise}{0.997}
\newcommand{\AccJellyRot}{0.756}
\newcommand{\AccJellyScale}{0.995}
\newcommand{\AccSintelBright}{1.000}
\newcommand{\AccSintelClean}{1.000}
\newcommand{\AccSintelCrfA}{1.000}
\newcommand{\AccSintelCrfB}{1.000}
\newcommand{\AccSintelCrfC}{1.000}
\newcommand{\AccSintelCropA}{0.502}
\newcommand{\AccSintelCropB}{0.589}
\newcommand{\AccSintelFps}{1.000}
\newcommand{\AccSintelNoise}{1.000}
\newcommand{\AccSintelRot}{0.797}
\newcommand{\AccSintelScale}{1.000}
\newcommand{\AttackCount}{11}

\newcommand{\BenchFrames}{240}

\newcommand{\BindCodeK}{207}
\newcommand{\BindCodeN}{255}
\newcommand{\BindCodeT}{6}

\newcommand{\ChromaN}{7}
\newcommand{\CorpusN}{1000}
\newcommand{\CorpusRatePct}{99.3}
\newcommand{\CorpusVerified}{993}
\newcommand{\CorpusWrongKeyTrials}{1000}
\newcommand{\CsBbbBright}{0.501}
\newcommand{\CsBbbClean}{0.498}
\newcommand{\CsBbbCrfA}{0.498}
\newcommand{\CsBbbCrfB}{0.495}
\newcommand{\CsBbbCrfC}{0.497}
\newcommand{\CsBbbCropA}{0.495}
\newcommand{\CsBbbCropB}{0.502}
\newcommand{\CsBbbFps}{0.493}
\newcommand{\CsBbbNoise}{0.500}
\newcommand{\CsBbbRot}{0.510}
\newcommand{\CsBbbScale}{0.495}
\newcommand{\CsJellyBright}{0.966}
\newcommand{\CsJellyClean}{0.954}
\newcommand{\CsJellyCrfA}{0.954}
\newcommand{\CsJellyCrfB}{0.601}
\newcommand{\CsJellyCrfC}{0.485}
\newcommand{\CsJellyCropA}{0.504}
\newcommand{\CsJellyCropB}{0.495}
\newcommand{\CsJellyFps}{0.960}
\newcommand{\CsJellyNoise}{0.652}
\newcommand{\CsJellyRot}{0.529}
\newcommand{\CsJellyScale}{0.954}

\newcommand{\CsSintelBright}{0.935}
\newcommand{\CsSintelClean}{0.938}
\newcommand{\CsSintelCrfA}{0.938}
\newcommand{\CsSintelCrfB}{0.857}
\newcommand{\CsSintelCrfC}{0.610}
\newcommand{\CsSintelCropA}{0.477}
\newcommand{\CsSintelCropB}{0.478}
\newcommand{\CsSintelFps}{0.942}
\newcommand{\CsSintelNoise}{0.476}
\newcommand{\CsSintelRot}{0.800}
\newcommand{\CsSintelScale}{0.939}
\newcommand{\FlickCsBbb}{+8.22}
\newcommand{\FlickCsJelly}{-0.34}
\newcommand{\FlickCsSintel}{+1.31}
\newcommand{\FlickDeltaMax}{-0.056}
\newcommand{\FlickDeltaMin}{-0.160}
\newcommand{\FlickOursBbb}{+0.08}
\newcommand{\FlickOursJelly}{-0.10}
\newcommand{\FlickOursSintel}{+0.04}
\newcommand{\GateAcceptPct}{71.3}
\newcommand{\GateAccepted}{189}
\newcommand{\GateMeasured}{265}
\newcommand{\GateScalePct}{97.4}
\newcommand{\IdxCropEightyVer}{2}
\newcommand{\IdxCropNinetyMax}{0.996}
\newcommand{\IdxCropNinetyMin}{0.990}
\newcommand{\IdxCropNinetyVer}{3}
\newcommand{\IdxFpsAccMax}{0.558}
\newcommand{\LiftAccepted}{112}
\newcommand{\LpipsMax}{0.0887}
\newcommand{\LpipsMin}{0.0335}
\newcommand{\MsssimMin}{0.9772}
\newcommand{\PayloadBits}{1024}
\newcommand{\PcBbbAFlick}{-0.143}
\newcommand{\PcBbbALpips}{0.0359}
\newcommand{\PcBbbAMsssim}{0.9772}
\newcommand{\PcBbbAPsnr}{35.52}
\newcommand{\PcBbbASsim}{0.9685}
\newcommand{\PcBbbBFlick}{-0.160}
\newcommand{\PcBbbBLpips}{0.0352}
\newcommand{\PcBbbBMsssim}{0.9774}
\newcommand{\PcBbbBPsnr}{35.52}
\newcommand{\PcBbbBSsim}{0.9687}
\newcommand{\PcJellyAFlick}{-0.154}
\newcommand{\PcJellyALpips}{0.0335}
\newcommand{\PcJellyAMsssim}{0.9826}
\newcommand{\PcJellyAPsnr}{38.23}
\newcommand{\PcJellyASsim}{0.9777}
\newcommand{\PcJellyBFlick}{-0.150}
\newcommand{\PcJellyBLpips}{0.0378}
\newcommand{\PcJellyBMsssim}{0.9821}
\newcommand{\PcJellyBPsnr}{38.29}
\newcommand{\PcJellyBSsim}{0.9760}
\newcommand{\PcSintelAFlick}{-0.072}
\newcommand{\PcSintelALpips}{0.0682}
\newcommand{\PcSintelAMsssim}{0.9825}
\newcommand{\PcSintelAPsnr}{40.79}
\newcommand{\PcSintelASsim}{0.9680}
\newcommand{\PcSintelBFlick}{-0.056}
\newcommand{\PcSintelBLpips}{0.0887}
\newcommand{\PcSintelBMsssim}{0.9825}
\newcommand{\PcSintelBPsnr}{40.79}
\newcommand{\PcSintelBSsim}{0.9657}

\newcommand{\PsnrBbb}{42.35}
\newcommand{\PsnrJelly}{43.48}
\newcommand{\PsnrMax}{40.79}
\newcommand{\PsnrMin}{35.52}
\newcommand{\PsnrSintel}{43.06}
\newcommand{\RecBbbAcc}{0.9990}

\newcommand{\RecJellyAcc}{0.9941}

\newcommand{\RecSintelAcc}{1.0000}

\newcommand{\ResidueCount}{7}
\newcommand{\RotClips}{2}

\newcommand{\RungDeepCount}{22}
\newcommand{\RungFortyCount}{176}
\newcommand{\RungGeqThirtyEightPct}{94.8}
\newcommand{\RungThirtyEightCount}{72}
\newcommand{\RungThirtySixCount}{23}
\newcommand{\RungTopCount}{700}
\newcommand{\SigCountBbb}{7}
\newcommand{\SigCountJelly}{7}
\newcommand{\SigCountSintel}{8}
\newcommand{\SsimMax}{0.9777}
\newcommand{\SsimMin}{0.9657}

\newcommand{\TransplantPairs}{30}
\newcommand{\VerBbbBright}{\ding{51}}
\newcommand{\VerBbbClean}{\ding{51}}
\newcommand{\VerBbbCrfA}{\ding{51}}
\newcommand{\VerBbbCrfB}{\ding{51}}
\newcommand{\VerBbbCrfC}{\ding{55}}
\newcommand{\VerBbbCropA}{\ding{55}}
\newcommand{\VerBbbCropB}{\ding{55}}
\newcommand{\VerBbbFps}{\ding{51}}
\newcommand{\VerBbbNoise}{\ding{51}}
\newcommand{\VerBbbRot}{\ding{55}}
\newcommand{\VerBbbScale}{\ding{51}}
\newcommand{\VerJellyBright}{\ding{51}}
\newcommand{\VerJellyClean}{\ding{51}}
\newcommand{\VerJellyCrfA}{\ding{51}}
\newcommand{\VerJellyCrfB}{\ding{51}}
\newcommand{\VerJellyCrfC}{\ding{55}}
\newcommand{\VerJellyCropA}{\ding{55}}
\newcommand{\VerJellyCropB}{\ding{55}}
\newcommand{\VerJellyFps}{\ding{51}}
\newcommand{\VerJellyNoise}{\ding{51}}
\newcommand{\VerJellyRot}{\ding{55}}
\newcommand{\VerJellyScale}{\ding{51}}
\newcommand{\VerSintelBright}{\ding{51}}
\newcommand{\VerSintelClean}{\ding{51}}
\newcommand{\VerSintelCrfA}{\ding{51}}
\newcommand{\VerSintelCrfB}{\ding{51}}
\newcommand{\VerSintelCrfC}{\ding{51}}
\newcommand{\VerSintelCropA}{\ding{55}}
\newcommand{\VerSintelCropB}{\ding{55}}
\newcommand{\VerSintelFps}{\ding{51}}
\newcommand{\VerSintelNoise}{\ding{51}}
\newcommand{\VerSintelRot}{\ding{55}}
\newcommand{\VerSintelScale}{\ding{51}}
\newcommand{\WrongKeyTrials}{33}

\title{Asymmetric Phase Coding Video Watermarking}

\author{Guang Yang\\
Phi Lab Foundation\\
{\tt\small guang.yang@philab.fund}
\and
Fengchen Liu\\
University of California, Berkeley\\
{\tt\small fengchenliu@berkeley.edu}
}

\begin{document}
\maketitle
\begin{abstract}
Existing video watermarking systems are symmetric: the party
that can verify a mark holds the extractor weights or generator secret and can
therefore also embed one. Benchmarks confirm the consequence, reporting that
white-box forgery defeats all evaluated methods. We present a training-free
video watermark that removes the shared secret. The signer embeds a complete
Ed25519 signature into the phase spectrum of the chroma plane; any party
holding the 32-byte public key and public per-video metadata verifies offline,
with no model, no registry, and no network. The payload, \PayloadBits{} bits
of signed message with error correction, is an order of magnitude above
common learned payloads and is carried by three design elements: a run-length
temporal layout whose decoder identifies payload groups by correlation and
never reads a frame index, a payload-free search that recovers scale,
rotation, and translation from the carrier itself, and a closed-loop signing
procedure that selects each video's embedding strength by self-verification
through the unchanged public verifier. On \CorpusN{} uncurated real-world
clips the system ships a verifying signature for \CorpusRatePct\% of the
corpus and accepts a wrong public key zero times in \CorpusWrongKeyTrials{}
attempts. An attack-aware acceptance gate yields embeddings that survive
H.264 re-encoding at 100\% and 50\% rescaling at \GateScalePct\% on gated
clips. The signature also verifies through a real display and capture loop,
an axis absent from published evaluations.
\end{abstract}
    
\section{Introduction}
\label{sec:intro}

\begin{figure*}[t!]
\centering
\includegraphics[width=\textwidth]{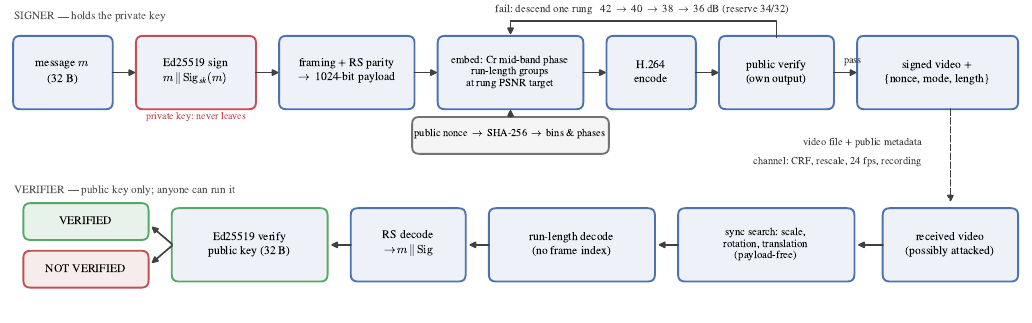}
\caption{System overview. The signer (top) signs the message with its
private Ed25519 key, frames it with Reed--Solomon parity into a
\PayloadBits{}-bit payload, and embeds it into mid-band Cr phase in
run-length groups; bin positions and phases derive from a public per-video
nonce. Signing is closed-loop: the encoded output is checked with the public
verifier, and on failure the signer descends one rung of the PSNR-target
ladder and re-embeds. The verifier (bottom) is fully public: payload-free
synchronization, index-free run-length decoding, Reed--Solomon decoding, and
the Ed25519 check yield \texttt{VERIFIED} or \texttt{NOT VERIFIED}. No
secret crosses the boundary between the two lanes.}
\label{fig:overview}
\end{figure*}

Video watermarking has made rapid progress on robustness. Learned systems
survive strong compression, rescaling, and editing, and ship in production
pipelines~\cite{fernandez2024videoseal,soucek2025pixelseal}. Yet nearly the
entire field shares one structural property: verification is symmetric.
Whoever can check a mark holds the extractor weights or the generator secret,
and that party could equally embed a mark of its own. Detection capability and
signing capability are the same capability. The consequence is measurable.
VideoMarkBench reports that white-box forgery succeeds against every video
watermarking method it evaluates, with both false-negative and false-positive
rates driven to one under small perturbations, and attributes this to
training that anticipates removal but not forgery~\cite{jiang2025videomarkbench}.
Recent theory reaches the matching conclusion from the other side: robustness
and public detectability are jointly hard to achieve, and schemes that publish
their detector give the attacker the removal oracle for
free~\cite{fairoze2025difficulty}.

Public-key cryptography resolved this exact problem for messages half a
century ago: signing is private, verification is public, and verification
grants no signing power. This paper carries that separation into video pixels.
The signer embeds a complete Ed25519 signature into the chroma phase spectrum
of the video. Any party holding the 32-byte public key, a public per-video
nonce, and the message length verifies offline. There is no extractor secret
to leak, because the extractor is public by design. Forging a mark that
verifies under the target key requires forging Ed25519, not probing a neural
detector~\cite{bernstein2012highspeed}.

The obstacle is capacity under distortion. A signature payload with framing
and error correction is \PayloadBits{} bits, an order of magnitude above the
96 to 128 bits common in learned systems~\cite{fernandez2024videoseal,
asnani2026flowmark}, and it must survive transform coding, frame-rate changes,
and rescaling without any learned decoder to absorb the damage. A wrong answer
is also worse than a missing answer: a signature that ``almost verifies'' does
not exist, so the transport layer must deliver the payload essentially intact
or the check fails. Four design choices meet these constraints, and each was
selected by measurement. First, the payload rides on mid-band phase of the Cr
chroma plane, where codecs spend few bits and the eye tolerates change; the
embedding amplitude is set from a per-frame distortion target. Second, a
run-length temporal layout holds each payload group for a run of consecutive
frames, and the decoder identifies groups by correlation without ever reading
a frame index, which is what survives frame-rate conversion and real screen
recording. Third, a payload-free synchronization search recovers scale,
rotation, and translation from the carrier itself before decoding. Fourth,
the distortion target is not fixed: a closed-loop signing procedure descends
a small ladder of targets and accepts the first embedding that verifies
through the public verifier, so every video is signed at its own operating
point.

We evaluate end to end, with the actual signature check as the acceptance
criterion and wrong-key controls in every run. On a controlled three-clip
benchmark with eleven attacks, the signature verifies through H.264 at
CRF~23 and 28, 50\% downscaling, 30 to 24\,fps conversion, noise, and
brightness change, with bit accuracy at or near 1.0, while a matched-payload
learned baseline decodes at chance on several of these axes.

\begin{figure*}[t!]
\centering
\includegraphics[width=\textwidth]{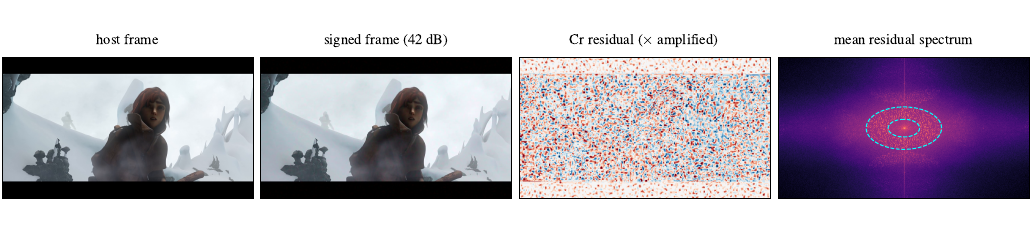}
\caption{A signed frame produced by the shipped implementation. The payload,
a \PayloadBits{}-bit Ed25519-signed message, modulates mid-band phase of the
Cr chroma plane; this clip accepted the mildest 42\,dB rung of the signing
ladder. The residual is
visually structureless, and its spectrum averaged over 48 frames concentrates
in the public carrier band of 0.05 to 0.12 cycles per pixel (dashed annulus),
where transform coding preserves it. Any holder of the 32-byte public key
verifies offline.}
\label{fig:pipeline}
\end{figure*}

On \CorpusN{}
uncurated real-world clips, closed-loop signing ships a verifying signature
for \CorpusVerified{} clips (\CorpusRatePct\%), and a wrong public key is
accepted zero times in \CorpusWrongKeyTrials{} attempts. An attack-aware
acceptance gate produces embeddings that survive H.264 re-encoding at 100\%
and rescaling at \GateScalePct\% on gated clips. The same configuration
survives a real display and capture loop through a video player and screen
recorder, an evaluation axis absent from prior video watermarking
evaluations.

Scope is stated as precisely as the claims. \texttt{VERIFIED} means the
extracted message carries a valid signature under the given key; failure means
the check could not complete and nothing more. The signature binds the message
bytes, not the surrounding pixels. We additionally evaluate, at small scale, a
fuzzy-commitment-style extension that associates the signed message with a
robust content feature~\cite{juels1999fuzzy,dodis2008fuzzy}, and we report its
tested scope explicitly. Visible geometric edits such as cropping are outside
the design threat model: an edit the eye can see does not need cryptography to
be noticed.

Our contributions are:
\begin{itemize}
  \item a training-free video watermark that embeds and publicly verifies real
        Ed25519 signatures; to the best of our knowledge, it is the first
        video watermarking system in which verification confers no
        embedding capability;
  \item a run-length temporal coding layer whose decoder never consults the
        received frame index, verified against frame-rate conversion where a
        frame-indexed layout fails at chance;
  \item closed-loop signing: because verification is public and
        self-contained, the signer selects each video's embedding strength by
        running the true acceptance test locally; on \CorpusN{} uncurated
        clips this ships verifying signatures for \CorpusRatePct\% of the
        corpus with zero wrong-key accepts, and an attack-aware gate turns
        attack survival into a signing-time decision;
  \item an end-to-end evaluation protocol whose acceptance criterion is the
        cryptographic check itself, including wrong-key negatives in every
        configuration, a corpus-scale study, and a real screen-recording
        loop.
\end{itemize}

\section{Related Work}
\label{sec:related}

\paragraph{Learned video watermarking.} Deep encoder-decoder systems descend
from image models such as HiDDeN and MBRS~\cite{zhu2018hidden,jia2021mbrs} and
now span two families. Post-hoc systems mark existing pixels: RivaGAN uses
attention to place bits~\cite{zhang2019rivagan}, DVMark embeds at multiple
scales~\cite{luo2021dvmark}, Video Seal scales training to production and
reaches 96 bits~\cite{fernandez2024videoseal}, Pixel Seal refines it with
adversarial-only objectives~\cite{soucek2025pixelseal}, FlowMark masks
embedding by motion~\cite{asnani2026flowmark}, DINVMark uses an invertible
network~\cite{ji2025dinvmark}, SPDMark displaces network
parameters~\cite{fares2025spdmark}, and ItoV adapts image models to video and
reports simulated screen-recording robustness~\cite{ye2023itov}. In-generation
systems mark the sampling process of a video diffusion model: VideoShield maps
watermark bits into the initial noise~\cite{hu2025videoshield}, and Stable
Signature and Gaussian Shading are the image
ancestors~\cite{fernandez2023stablesignature,yang2024gaussianshading}, with
VidStamp, LVMark, VideoMark, Video Signature, and SIGMark extending capacity
and blind extraction~\cite{teymoorianfard2025vidstamp,jang2024lvmark,
hu2025videomark,huang2025vidsig,sigmark2026}. Payloads range from 32 to a few
hundred bits in the post-hoc family; SIGMark reaches 8192 bits but requires
the generation-side latent pathway. In every one of these systems the
verifier's artifact, extractor weights or generator secret, suffices to embed,
so verification implies signing capability. Our system differs at that root:
verification uses a public key that confers no embedding power, and the
payload is a real signature rather than an identifier looked up elsewhere.

\paragraph{Attacks and benchmarks.} WAVES standardizes image watermark
stress-testing~\cite{an2024waves}, and VideoMarkBench extends it to video,
finding that forgery, not removal, is the undefended
direction~\cite{jiang2025videomarkbench}. Regeneration attacks remove
invisible marks with provable guarantees by re-synthesizing the
content~\cite{zhao2024regeneration}; such attacks erase marks but cannot
create a signature that verifies under someone else's key, which is the
property this paper adds. The SoK of watermarking for AI-generated content
frames these threat models systematically~\cite{zhao2024sok}.

\paragraph{Asymmetric and publicly verifiable watermarking.} The goal of
detaching detection from embedding is old. Furon and Duhamel proposed
asymmetric detection where the detector does not hold the embedding
key~\cite{furon1999asymmetric}, and Eggers, Su, and Girod built public
detection from eigenvectors of linear transforms, noting themselves that
published detectors invite targeted removal~\cite{eggers2000publickey}.
Hopper, Molnar, and Wagner formalized strong watermarking against arbitrary
channels~\cite{hopper2007weak}, and recent analysis shows robustness plus
public detectability remains fundamentally difficult when the detector must
also resist removal~\cite{fairoze2025difficulty}. Our position differs from
classical asymmetric detection in what is public: we publish the entire
decoding path and rest security only on signature unforgeability, accepting
that an adversary who sees the decoder can attempt removal, which is the
visible-damage regime the threat model already concedes. In language models,
publicly detectable and publicly verifiable schemes exist with cryptographic
guarantees~\cite{fairoze2023publicly,liu2024unforgeable}, and Puppy provides
public verifiability through interactive protocols~\cite{isler2024puppy};
none of these transports a signature through video pixels.

\paragraph{Copy attacks and content binding.} Kutter, Voloshynovskiy, and
Herrigel showed that a watermark estimated from one asset can be re-embedded
into another~\cite{kutter2000copyattack}, and the template attack removes
synchronization patterns~\cite{herrigel2001template}. Deguillaume et
al.\ answered with hybrid schemes that bind the robust mark to content through
a fragile layer~\cite{deguillaume2002hybrid,deguillaume2003hybrid}, and
Adelsbach, Katzenbeisser, and Veith gave provable constructions against copy
and ambiguity attacks~\cite{adelsbach2003provably}. Our optional binding layer
follows the fuzzy commitment and secure sketch line of Juels and Wattenberg
and Dodis et al.~\cite{juels1999fuzzy,dodis2008fuzzy,juels2006fuzzyvault,
boyen2004reusable}: a public helper string lets the verifier reproduce a
content digest from received pixels, and the digest sits inside the signed
message. We evaluate this extension at small scale and state its limits in
Section~\ref{sec:exp-binding}.

\paragraph{Phase-domain embedding.} Phase carriers for video are nearly
unexplored. Meenakshi et al.\ modulate DFT phase per frame with BPSK but
report no modern codec results~\cite{meenakshi2014phase}; classical
codec-robust designs stay in magnitude or transform-coefficient
domains~\cite{zarmehi2017robust,elrowayati2020hevc}. PhaseMark, concurrent
work on images, injects phase patterns in the latent frequency domain of a
generator's autoencoder~\cite{lee2026phasemark}. To our knowledge no prior
system transports a cryptographic signature on video phase through H.264,
frame-rate conversion, and screen recording.

\section{Method}
\label{sec:method}

Figure~\ref{fig:overview} shows the full system: a signing lane that holds
the private key and runs the closed loop, and a public verification lane
that anyone can execute. This section specifies each stage.

\subsection{Threat model and verification semantics}
\label{sec:threat}

We separate two roles that symmetric watermarking merges. The \emph{signer}
holds an Ed25519 private key. The \emph{verifier} holds the 32-byte public key
and public decoding metadata: a per-video nonce, the embedding mode, and the
message length. The verifier runs offline. It needs no shared secret, no model
weights, no original video, no registry, and no network service.

The system covers transformations that preserve the visible content of the
frame: re-encoding, frame-rate conversion, rescaling, additive noise, and
brightness change. Visible geometric edits such as cropping are out of scope by
design. A verifier that tolerated a 20\% crop would certify a frame missing one
fifth of its content; for evidentiary use the correct answer to a visibly
edited video is \texttt{NOT VERIFIED}, and that is the answer the system
gives. Verification returns one of two answers. \texttt{VERIFIED} states that
the extracted payload carries a valid signature under the given public key.
\texttt{NOT VERIFIED} states that the check could not be completed. A failed
check is not evidence of forgery. It can result from heavy degradation, a wrong
key, or an unmarked video. This asymmetry is deliberate. The verifier never
promotes a damaged decode into a positive answer, because the signature check
either passes exactly or fails.

Security rests on the signature, not on secrecy of the carrier. The layout
nonce is public. Knowing it allows decoding. It does not allow signing, because
creating a payload that verifies under the target public key requires the
private key. Unforgeability is the standard computational property of Ed25519
under correct implementation~\cite{bernstein2012highspeed,rfc8032}.

\subsection{Payload construction}
\label{sec:payload}

The payload frames a message $m$ of known length with its signature:
\begin{equation}
  b \;=\; \mathrm{flag} \,\|\, \mathrm{len}(m) \,\|\, m \,\|\,
  \mathrm{Sig}_{sk}(m),
  \label{eq:payload}
\end{equation}
where $\mathrm{Sig}_{sk}(m)$ is the 64-byte Ed25519 signature. A Reed--Solomon
code with 30 parity bytes protects the framed bytes against residual bit
errors. For the 32-byte messages used in our experiments the coded payload is
\PayloadBits{} bits. This is an order of magnitude above common neural payload
sizes, which sit near 96 to 128 bits~\cite{fernandez2024videoseal,
asnani2026flowmark}, and the signature is the reason: a real signature cannot
be shortened below 64 bytes, so a signature-carrying watermark must solve the
capacity problem, not avoid it.

\subsection{Chroma phase carrier}
\label{sec:carrier}

Each payload bit modulates the phase of a set of mid-band frequency bins of the
Cr chroma plane. A radial band of 0.05 to 0.12 cycles per pixel carries all
slots. The band is low enough to survive transform coding and high enough to
avoid the DC region where the eye tracks color casts. We embed in chroma
because H.264 and H.265 allocate fewer bits there, and because luma phase
carries edge structure that the eye notices first. The amplitude of each
modification is set per frame from a target embedding distortion expressed as
PSNR, so quality is a controlled input rather than an outcome; the target
itself is selected per video by the closed loop of
Section~\ref{sec:closedloop}.

The bin positions and carrier phases for every slot derive from a public
per-video nonce through SHA-256. The nonce provides domain separation: two
videos signed by the same key place their carriers on different bins, so
residual averaging across a collection of signed videos does not accumulate a
common pattern. The nonce travels as plain metadata next to the file, in the
same way a provenance manifest travels with an asset.

\subsection{Run-length temporal coding}
\label{sec:temporal}

\begin{figure}[t!]
\centering
\includegraphics[width=\columnwidth]{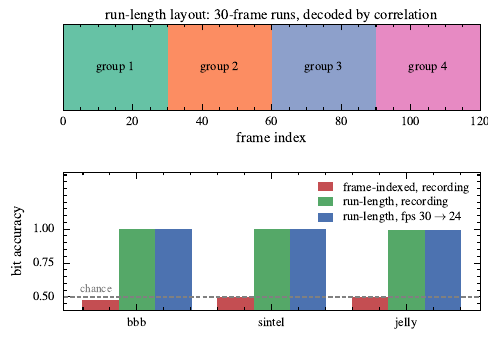}
\caption{Top: the run-length layout cycles \(G=4\) payload groups in 30-frame
runs; the decoder assigns each received frame to a group by correlation and
never reads a frame index. Bottom: measured consequence. After real screen
recording, a frame-indexed layout decodes at chance while run-length recovers
the payload on all three clips; frame-rate conversion behaves the same way.}
\label{fig:temporal}
\end{figure}

A layout keyed to the frame index fails whenever frame timing changes, because
frame $t$ after conversion is not frame $t$ before it. Our decoder therefore
never consults the received frame index. The payload is split into $G=4$
groups. Each group occupies $K=30$ consecutive frames, and the cycle repeats
for the duration of the video. Every frame is marked. At the decoder, each
received frame is correlated against the four candidate group layouts, and the
best-matching group claims the frame. Soft evidence accumulates per group, the
groups concatenate into the coded payload, and Reed--Solomon decoding plus the
signature check complete verification.

Group identity is recoverable from content alone because the four layouts are
orthogonal by construction: they draw disjoint bin subsets from the carrier
band. Frame drops, duplications, and rate conversion shift which frames land in
a group but do not change which group a frame belongs to. Section
\ref{sec:exp-temporal} measures this property directly against a frame-indexed
alternative on identical clips.

\subsection{Blind geometric synchronization}
\label{sec:sync}

Phase carriers are sensitive to coordinate changes: a rescale moves every bin,
and a translation rotates every phase in proportion to its frequency. Before
decoding, the verifier runs a payload-free search over scale, rotation, and
translation. The search maximizes carrier energy at the nonce-derived bin
positions, using the embedded carrier itself as the alignment target rather
than a separate template. Local spectral whitening precedes the search, which
flattens host energy so that textured clips do not dominate the objective. The
recovered transform is inverted and decoding proceeds on the aligned frames.
The alignment stage reports its estimate, which makes failures visible: a
wrong scale estimate produces a near-chance bit accuracy rather than a silent
wrong answer, and the signature gate converts near-chance decodes into
\texttt{NOT VERIFIED}.

\subsection{Closed-loop signing}
\label{sec:closedloop}

No single embedding strength suits all content. The host's own energy inside
the carrier band varies by more than an order of magnitude across real
footage and acts as a noise floor for the carrier
(Section~\ref{sec:exp-diagnostics}), so a target that is generous on flat
content is insufficient on saturated, textured content. The signer therefore
selects the operating point per video. It embeds at the mildest target of a
fixed ladder of per-frame PSNR targets (42, 40, 38, 36\,dB, with reserve
rungs at 34 and 32\,dB), runs the public verifier on the encoded result, and
descends one rung on failure. The first rung that verifies is shipped, along
with the same public metadata as before. The acceptance test is the exact
verification of Section~\ref{sec:threat} under the signer's own public key;
the verifier binary never changes, and the shipped file carries no record of
the search.

The acceptance check extends to attacks. An attack-aware gate re-encodes the
candidate under an attack of interest, for example conversion to 24\,fps, and
accepts a rung only when both the clean file and the attacked file verify.
Because the gate discards low-margin embeddings, the clips it accepts survive
other attacks at high rates (Section~\ref{sec:exp-corpus}).

Closed-loop signing is possible precisely because verification is public and
self-contained. The signer runs the true acceptance test locally, with no
oracle problem and no held-out secret. The cost is one sign-and-verify pass
per rung, paid once at signing time. Verification cost and semantics are
unchanged.

\subsection{What the signature does and does not bind}
\label{sec:binding}

The signature binds the message bytes. Any party can therefore check that the
message was signed by the key holder and has not been altered. The signature
by itself does not bind the surrounding pixels: in the classical watermark
copy attack~\cite{kutter2000copyattack}, an adversary estimates the mark
carried by one asset and re-embeds it into another, transplanting a valid
signature onto footage the signer never saw. Where
deployment requires tying the signed statement to the carrying content, the
payload can additionally carry a public helper string that associates the
message with a robust content feature in the spirit of fuzzy
commitment~\cite{juels1999fuzzy,dodis2008fuzzy}. We evaluate such an extension
at small scale in Section~\ref{sec:exp-binding} and state its scope there.
Characterizing that association against a searching adversary at corpus scale
remains open, and none of the claims in this paper depend on it.
\section{Experiments}
\label{sec:experiments}

\subsection{Setup}
\label{sec:setup}

\paragraph{Implementation.} The full system is a single training-free program.
Signing embeds the payload of Section~\ref{sec:payload} into the Cr plane at a
per-frame PSNR target chosen by the closed loop of
Section~\ref{sec:closedloop} and re-encodes with H.264 at CRF~23.
Verification decodes with the public key, the public nonce, and the message
length. No stage contains a learned component, and one verifier binary
produces every number in this section.

\paragraph{Controlled benchmark.} Mechanism-level results use three
1280$\times$720 clips of \BenchFrames{} frames each, drawn from open cinematic
and photographic test material and re-encoded at high quality as sources. The
payload is \PayloadBits{} bits: a 32-byte message, the 64-byte Ed25519
signature, framing, and Reed--Solomon parity. All attacks run through
\texttt{ffmpeg} on the encoded output. Every table cell states bit accuracy in
the extracted coded payload and whether the Ed25519 signature verified. A
wrong-key control runs in every configuration: across all \WrongKeyTrials{}
clip and attack combinations, a mismatched public key was accepted zero times.

\paragraph{Corpus-scale study.} System-level results use \CorpusN{} uncurated
UHD clips from Inter4K~\cite{stergiou2021adapool}, downscaled to 720p at 120
frames, disjoint from the three development clips. Each clip is signed with a
unique public nonce and verified end to end through the CLI; every signed
clip is additionally checked against a mismatched public key. Rotation search
is disabled in this protocol because no rotation attack is applied; rotation
evidence comes from the controlled benchmark.

\paragraph{Baseline.} We compare against ChunkySeal, a re-trained
high-capacity variant of the VideoSeal architecture~\cite{fernandez2024videoseal},
configured to the same 1024-bit payload on the same clips and attacks. Its
in-memory round trip reaches 0.995 to 1.000 bit accuracy on these clips, so
the comparison starts from a healthy operating point. The two systems answer
different questions: ChunkySeal detects with a learned private extractor,
while our verifier is public. We therefore report per-axis numbers and no
aggregate winner.

\subsection{Robustness of the shipped configuration}
\label{sec:exp-main}

\begin{figure*}[t!]
\centering
\includegraphics[width=\textwidth]{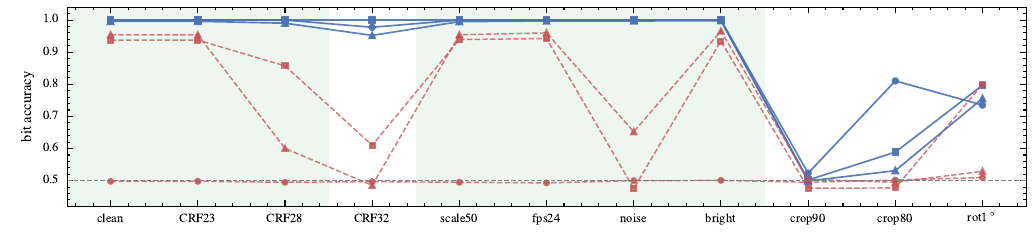}
\caption{Bit accuracy per attack on the three benchmark clips: ours (solid
blue) against ChunkySeal at the same \PayloadBits{}-bit payload (dashed red);
markers denote clips. Shaded columns are attacks where the Ed25519 signature
verified on 3/3 clips. The dashed grey line is chance. Crop and plain rotation
are outside the design scope; rotation is restored by the synchronization
search of Section~\ref{sec:sync}.}
\label{fig:robustness}
\end{figure*}

\begin{figure}[t!]
\centering
\includegraphics[width=\columnwidth]{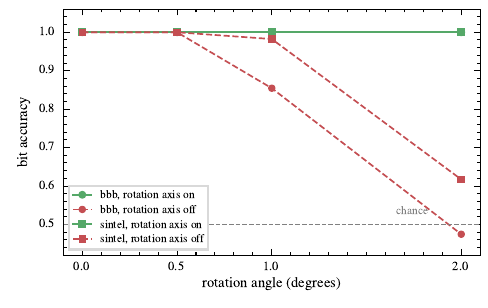}
\caption{The rotation axis of the synchronization search, isolated on
\RotClips{} benchmark clips. With the axis enabled (solid), the payload-free
search recovers the angle and bit accuracy stays at 1.000 through 2 degrees;
with the axis disabled (dashed), accuracy collapses to chance. The signature
verified in every solid-line cell and no wrong key was accepted.}
\label{fig:rotation}
\end{figure}

Table~\ref{tab:main} reports the benchmark, and Figure~\ref{fig:robustness}
plots the same measurements. Bit accuracy is 1.000 or within
rounding of it for H.264 CRF~23 and CRF~28, 50\% downscale with restoration,
frame-rate conversion from 30 to 24\,fps, additive noise, and brightness
change, on all three clips, and the signature verifies in each of these cells.
CRF~32 verifies on one clip of three: compression noise at that level begins
to exceed the Reed--Solomon budget on textured content. Cropping and rotation
without resynchronization sit at or near chance, which matches the design
scope of Section~\ref{sec:threat}; rotation becomes recoverable when the
synchronization search of Section~\ref{sec:sync} includes the rotation axis.
Figure~\ref{fig:rotation} isolates that axis: with the axis enabled, bit
accuracy is 1.000 at every tested angle up to 2 degrees on both measured
clips, while the same decodes without the axis fall from 1.000 to chance.

Crop coverage exists elsewhere in the design space, at a measured price. A
frame-indexed layout variant, run on the same clips and payload, anchors its
carrier to the absolute frame grid, and the synchronization search then
recovers the crop window: it verifies after cropping to 90\% on
\IdxCropNinetyVer{}/3 clips (bit accuracy \IdxCropNinetyMin{} to
\IdxCropNinetyMax{}) and after cropping to 80\% on \IdxCropEightyVer{}/3. The
same variant decodes frame-rate conversion at chance (\IdxFpsAccMax{} at
best). Spatial anchoring and temporal robustness trade against each other;
the shipped default keeps the temporal axis that
Sections~\ref{sec:exp-temporal} and \ref{sec:exp-recapture} measure, and
whether a verifier should accept a visibly cropped frame at all is the policy
question of Section~\ref{sec:threat}.

\begin{table*}[t!]
\centering
\small
\begin{tabular}{l ccc ccc ccc}
\toprule
& \multicolumn{3}{c}{bbb} & \multicolumn{3}{c}{sintel} & \multicolumn{3}{c}{jellyfish} \\
\cmidrule(lr){2-4}\cmidrule(lr){5-7}\cmidrule(lr){8-10}
Attack & Acc & Sig & CS & Acc & Sig & CS & Acc & Sig & CS \\
\midrule
clean            & \AccBbbClean  & \VerBbbClean  & \CsBbbClean  & \AccSintelClean  & \VerSintelClean  & \CsSintelClean  & \AccJellyClean  & \VerJellyClean  & \CsJellyClean \\
H.264 CRF 23     & \AccBbbCrfA   & \VerBbbCrfA   & \CsBbbCrfA   & \AccSintelCrfA   & \VerSintelCrfA   & \CsSintelCrfA   & \AccJellyCrfA   & \VerJellyCrfA   & \CsJellyCrfA \\
H.264 CRF 28     & \AccBbbCrfB   & \VerBbbCrfB   & \CsBbbCrfB   & \AccSintelCrfB   & \VerSintelCrfB   & \CsSintelCrfB   & \AccJellyCrfB   & \VerJellyCrfB   & \CsJellyCrfB \\
H.264 CRF 32     & \AccBbbCrfC   & \VerBbbCrfC   & \CsBbbCrfC   & \AccSintelCrfC   & \VerSintelCrfC   & \CsSintelCrfC   & \AccJellyCrfC   & \VerJellyCrfC   & \CsJellyCrfC \\
scale 50\%       & \AccBbbScale  & \VerBbbScale  & \CsBbbScale  & \AccSintelScale  & \VerSintelScale  & \CsSintelScale  & \AccJellyScale  & \VerJellyScale  & \CsJellyScale \\
fps 30$\to$24    & \AccBbbFps    & \VerBbbFps    & \CsBbbFps    & \AccSintelFps    & \VerSintelFps    & \CsSintelFps    & \AccJellyFps    & \VerJellyFps    & \CsJellyFps \\
Gaussian noise   & \AccBbbNoise  & \VerBbbNoise  & \CsBbbNoise  & \AccSintelNoise  & \VerSintelNoise  & \CsSintelNoise  & \AccJellyNoise  & \VerJellyNoise  & \CsJellyNoise \\
brightness       & \AccBbbBright & \VerBbbBright & \CsBbbBright & \AccSintelBright & \VerSintelBright & \CsSintelBright & \AccJellyBright & \VerJellyBright & \CsJellyBright \\
crop 90\%        & \AccBbbCropA  & \VerBbbCropA  & \CsBbbCropA  & \AccSintelCropA  & \VerSintelCropA  & \CsSintelCropA  & \AccJellyCropA  & \VerJellyCropA  & \CsJellyCropA \\
crop 80\%        & \AccBbbCropB  & \VerBbbCropB  & \CsBbbCropB  & \AccSintelCropB  & \VerSintelCropB  & \CsSintelCropB  & \AccJellyCropB  & \VerJellyCropB  & \CsJellyCropB \\
rotate 1$^\circ$ & \AccBbbRot    & \VerBbbRot    & \CsBbbRot    & \AccSintelRot    & \VerSintelRot    & \CsSintelRot    & \AccJellyRot    & \VerJellyRot    & \CsJellyRot \\
\midrule
signature verified & \multicolumn{3}{c}{\SigCountBbb/\AttackCount} & \multicolumn{3}{c}{\SigCountSintel/\AttackCount} & \multicolumn{3}{c}{\SigCountJelly/\AttackCount} \\
\bottomrule
\end{tabular}
\caption{Bit accuracy (Acc), Ed25519 verification (Sig), and ChunkySeal bit
accuracy (CS) under identical attacks, three 720p clips, \PayloadBits{}-bit
payload for both systems. Crop and plain rotation are outside the design scope
(Section~\ref{sec:threat}); rotation is restored by the synchronization search
when its rotation axis is enabled. A wrong public key was accepted in
0 of \WrongKeyTrials{} runs.}
\label{tab:main}
\end{table*}

Embedding quality on these clips is \PsnrBbb{}, \PsnrSintel{}, and
\PsnrJelly{}\,dB PSNR against the unmarked source. Temporal flicker, measured
as the mean absolute inter-frame difference relative to the host, changes by
\FlickOursBbb{}\%, \FlickOursSintel{}\%, and \FlickOursJelly{}\%, against
\FlickCsBbb{}\%, \FlickCsSintel{}\%, and \FlickCsJelly{}\% for the baseline on
the same clips.

\subsection{Temporal layout ablation}
\label{sec:exp-temporal}

The run-length decoder of Section~\ref{sec:temporal} is the reason frame-rate
conversion in Table~\ref{tab:main} verifies at accuracy near 1.0. A
frame-indexed layout measured on the same three clips decodes at chance after
30 to 24\,fps conversion, because the received frame index no longer matches
the embedding index. The same contrast appears end to end in the screen
recording study below, where the frame-indexed variant fails on all three
clips while run-length verifies on all three.

\subsection{Screen recording}
\label{sec:exp-recapture}

We evaluate a full display and capture loop: the signed video plays in a real
video player on a virtual X11 display, a screen recorder captures the display
at its own clock, and the capture is re-encoded. The recorded streams gain
frames relative to the source (255 recorded for 240 played), shift alignment
by roughly 35 frames, and lose 4 to 12\,dB PSNR. The run-length configuration
verifies on all three clips at bit accuracies \RecBbbAcc{}, \RecSintelAcc{},
and \RecJellyAcc{}. Unmarked recordings are rejected in every case. To our
knowledge, published video watermarking evaluations do not include a real
display and capture loop; the closest reported axis is software-simulated
recording~\cite{ye2023itov}.

\subsection{Perceptual cost}
\label{sec:exp-quality}

\begin{table}[t!]
\centering
\small
\setlength{\tabcolsep}{4.5pt}
\begin{tabular}{l ccccc}
\toprule
Clip & PSNR & SSIM & MS-SSIM & LPIPS & $\Delta$flicker \\
\midrule
sintel-a & \PcSintelAPsnr & \PcSintelASsim & \PcSintelAMsssim & \PcSintelALpips & \PcSintelAFlick \\
sintel-b & \PcSintelBPsnr & \PcSintelBSsim & \PcSintelBMsssim & \PcSintelBLpips & \PcSintelBFlick \\
bbb-a    & \PcBbbAPsnr    & \PcBbbASsim    & \PcBbbAMsssim    & \PcBbbALpips    & \PcBbbAFlick \\
bbb-b    & \PcBbbBPsnr    & \PcBbbBSsim    & \PcBbbBMsssim    & \PcBbbBLpips    & \PcBbbBFlick \\
jelly-a  & \PcJellyAPsnr  & \PcJellyASsim  & \PcJellyAMsssim  & \PcJellyALpips  & \PcJellyAFlick \\
jelly-b  & \PcJellyBPsnr  & \PcJellyBSsim  & \PcJellyBMsssim  & \PcJellyBLpips  & \PcJellyBFlick \\
\bottomrule
\end{tabular}
\caption{Embedding cost at the mildest rung on six 720p clips: PSNR (dB),
SSIM, MS-SSIM, LPIPS against the unmarked source, and the change in
inter-frame flicker (mean absolute inter-frame difference; negative means
the signed video flickers less than its host).}
\label{tab:quality}
\end{table}

Embedding distortion is a controlled input: each accepted embedding meets its
rung's per-frame PSNR target by construction, and
Figure~\ref{fig:ladder} reports exactly which rung every corpus clip
received, so the quality operating point of each signed video is known
rather than estimated. Table~\ref{tab:quality} reports full-reference
metrics at the mildest rung on six 720p clips of 96 frames: \PsnrMin{} to
\PsnrMax{}\,dB PSNR, \SsimMin{} to \SsimMax{} SSIM~\cite{wang2004ssim},
MS-SSIM at least \MsssimMin{}, and \LpipsMin{} to \LpipsMax{}
LPIPS~\cite{zhang2018lpips}. Inter-frame flicker decreases on every clip, by
\FlickDeltaMax{} to \FlickDeltaMin{} absolute difference units, because the
added carrier is temporally coherent within each group.

\subsection{Content association at small scale}
\label{sec:exp-binding}

\begin{figure}[t!]
\centering
\includegraphics[width=\columnwidth]{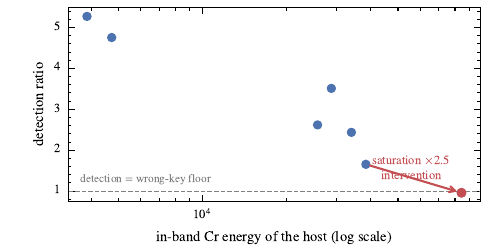}
\caption{Host in-band Cr energy against the detection ratio on the
\ChromaN{}-clip panel: more host energy inside the carrier band means a worse
ratio, so the host is a noise floor, not a carrier. Red: raising one clip's
saturation moved its energy 2.2$\times$ and pushed detection to the
wrong-key floor.}
\label{fig:chroma}
\end{figure}

The optional binding extension of Section~\ref{sec:binding} carries a public
helper string built from a BCH$[\BindCodeN,\BindCodeK]$ code with correction
radius $t=\BindCodeT$ over a 255-bit temporally pooled DCT feature. On a
six-clip corpus, the association behaves as designed: the feature survives all
seven non-geometric channels with zero decoded-feature errors across 42
measurements, re-verification succeeds on all six clips, and all
\TransplantPairs{} ordered cross-video payload transfers at the protocol level
are rejected, because the recovered feature of the receiving video disagrees
with the digest inside the signed message. We report this as a protocol-level
result on the tested corpus. Its behavior against a searching adversary over a
large corpus is not established here, and the claims of
Sections~\ref{sec:exp-main} through \ref{sec:exp-quality} do not rest on it.

\subsection{Corpus-scale verification}
\label{sec:exp-corpus}

\begin{figure}[t!]
\centering
\includegraphics[width=\columnwidth]{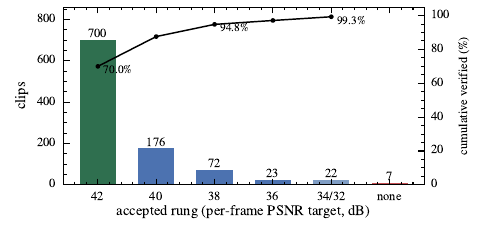}
\caption{Closed-loop signing over \CorpusN{} uncurated Inter4K clips: clips
per accepted rung (bars) and cumulative verified fraction (line).
Descending the ladder ships verifying signatures for \CorpusVerified{} clips
(\CorpusRatePct\%); a wrong public key was accepted zero times in
\CorpusWrongKeyTrials{} attempts.}
\label{fig:ladder}
\end{figure}

Closed-loop signing is evaluated on all \CorpusN{} corpus clips end to end
through the shipped CLI. Figure~\ref{fig:ladder} shows the outcome as a rung
distribution. \RungTopCount{} clips accept at the mildest 42\,dB target;
\RungFortyCount{}, \RungThirtyEightCount{}, and \RungThirtySixCount{} accept
at 40, 38, and 36\,dB; the reserve rungs recover \RungDeepCount{} more. In
total \CorpusVerified{} of \CorpusN{} clips (\CorpusRatePct\%) ship a
verifying signature, \RungGeqThirtyEightPct\% of the corpus at 38\,dB or
milder. The remaining \ResidueCount{} clips are the measured boundary of the
phase carrier at the evaluated targets. Across all \CorpusWrongKeyTrials{}
wrong-key verifications on this corpus, zero were accepted, and the cumulative
wrong-key count across every experiment in this paper is zero.

The attack-aware gate of Section~\ref{sec:closedloop} was evaluated with
24\,fps conversion as the gating attack on the clips that reject the mildest
rung, the hardest subpopulation of the corpus. The gate accepted
\GateAccepted{} of \GateMeasured{} measured clips (\GateAcceptPct\%) at a rung
whose attacked copy also verifies. On these gated clips, survival under the
non-gating attacks is 100\% for H.264 CRF~23 re-encoding and \GateScalePct\%
for 50\% rescaling: selecting for margin under one attack yields embeddings
robust under others. On a seed-fixed sample of \LiftAccepted{} corpus clips
whose mildest-rung embedding verifies clean but fails 24\,fps conversion, the
gate found an accepting rung for every sampled clip, making attack survival a
signing-time decision rather than a fixed property of the content.

\subsection{Carrier diagnostics}
\label{sec:exp-diagnostics}

Two observed regularities motivate the closed loop and make deployment
predictable. First, on a \ChromaN{}-clip panel, the host's own energy inside
the carrier band acts as a noise floor: clips with more in-band Cr energy
detect worse, with a strong negative rank correlation, and raising one clip's
saturation to move its in-band energy up by a factor of 2.2 moved its
detection ratio from healthy to failing (Figure~\ref{fig:chroma}). Because
this noise floor spans more than an order of magnitude across real footage,
no single embedding strength serves all content, which is exactly the gap the
per-video ladder of Section~\ref{sec:closedloop} closes. Second,
synchronization quality improves with resolution: identity recovery rises
monotonically from 360p to 1080p in a four-resolution sweep, consistent with
the number of usable carrier bins growing with frame area. Both are
diagnostics on the tested clips rather than fitted laws.

\section{Discussion and Conclusion}
\label{sec:conclusion}

\paragraph{What is established.} A training-free watermark can carry a
complete Ed25519 signature through the transformations that dominate real
video distribution: transform coding, rescaling, frame-rate conversion, noise,
brightness change, and a real display and capture loop. Verification needs
only the public key and public metadata, so detection capability no longer
implies signing capability. At corpus scale, closed-loop signing ships a
verifying signature for \CorpusRatePct\% of \CorpusN{} uncurated real-world
clips, with the accepted quality target of every clip known by construction.
The acceptance criterion throughout is the signature check itself; wrong-key
controls were accepted zero times in every tested configuration, including
all \CorpusWrongKeyTrials{} corpus-scale attempts, and degraded inputs
produce \texttt{NOT VERIFIED}, never a false positive.

\paragraph{Deployment.} The verifier needs three public items next to the
file: the nonce, the mode, and the message length. These travel like any
provenance metadata, for example inside a sidecar manifest. Publishing the
decoder is part of the design: security rests on the private key, and the
public nonce only separates carrier layouts across videos. The closed loop
concentrates all adaptivity at the signer, where the private key already
lives: signing pays one sign-and-verify pass per rung, the attack-aware gate
turns robustness targets into signing-time policy, and the verifier binary
is identical for every video regardless of rung.

\paragraph{Limitations.} Visible geometric edits are out of scope by design:
cropping breaks decoding, and the system reports failure rather than tolerance
of a visibly altered frame. A small residue of the corpus
(\ResidueCount{} of \CorpusN{} clips) rejects every evaluated rung and marks
the measured boundary of the phase carrier. The signature binds the message
bytes; the content-association extension that ties the message to a robust
feature is validated on a six-clip corpus, and its behavior against a
searching adversary at large corpus scale is an open problem we state rather
than claim. Camera recapture of physical displays and platform round-trips
remain untested.

\paragraph{Outlook.} The capacity, corpus-scale coverage, and closed-loop
operating points demonstrated here make signature transport a usable
primitive for video; per-video content certification at corpus scale and
physical-display recapture are the natural next steps.

{
    \small
    \bibliographystyle{ieeenat_fullname}
    \bibliography{main}
}

\end{document}